\documentclass[sigconf,nonacm]{acmart}

\renewcommand\footnotetextcopyrightpermission[1]{}

\usepackage{booktabs}
\usepackage{graphicx}
\usepackage{amsmath}

\usepackage{amssymb}

\usepackage{mdframed}

\begin{document}

\title{ReasonOps: A Unified Operational Paradigm for Trustworthy Verified LLM Reasoning}

\author{Adnan Rashid}
\affiliation{
  \institution{School of Electrical Engineering and Computer Science (SEECS)\\ National University of Sciences and Technology (NUST)}
  \country{Islamabad, Pakistan}
}
\email{adnan.rashid@seecs.edu.pk}

\begin{abstract}
Large Language Models (LLMs) have transformed artificial intelligence from primarily generative systems into increasingly capable reasoning agents. Recent advances in theorem proving, autoformalization, symbolic reasoning, and tool-augmented language models demonstrate substantial progress toward machine-assisted formal reasoning. However, current reasoning systems still suffer from hidden logical inconsistencies, hallucinated symbolic transitions, unsupported theorem applications, and limited reliability guarantees. Existing approaches remain fragmented across formal verification, runtime assurance, neuro-symbolic reasoning and trustworthy Artificial Intelligence (AI) research communities.

This paper introduces \textit{ReasonOps}, a unified operational paradigm for trustworthy verified reasoning systems. Inspired by operational ecosystems such as DevOps and MLOps, ReasonOps treats reasoning as a continuously monitored, verifiable, reliability-aware operational process rather than an isolated inference task. The proposed paradigm integrates semantic interpretation, autoformalization, symbolic reasoning, theorem proving, runtime assurance, probabilistic reliability estimation, and adaptive correction into a unified reasoning lifecycle. The paper further presents the ReasonOps architecture, demonstrates its workflow using an autonomous braking system analysis example, and discusses its potential role in future safety-critical autonomous AI systems. We argue that operational reasoning paradigms such as ReasonOps may become foundational infrastructure for next-generation trustworthy AI ecosystems.
\end{abstract}

\keywords{Verified Reasoning, Autoformalization, Formal
Verification, Symbolic Reasoning, Runtime Assurance, Trustworthy AI}

\maketitle

\section{Introduction}

The rapid emergence of Large Language Models (LLMs) has fundamentally reshaped the landscape of artificial intelligence. Modern reasoning-capable systems can now solve mathematical problems, synthesize programs, interact with theorem provers, generate formal proofs, retrieve symbolic knowledge, and perform increasingly sophisticated multi-step reasoning tasks. These developments have accelerated research in autoformalization, theorem proving, and neuro-symbolic Artificial Intelligence (AI) while simultaneously raising important questions concerning reasoning reliability and trustworthiness \cite{yang2023leandojo,azerbayev2023llemma,polu2020generative}.

Recent advances in retrieval-augmented theorem proving, formal mathematical reasoning benchmarks, and proof synthesis systems have demonstrated that LLMs can increasingly cooperate with symbolic reasoning engines such as Lean, Coq, Isabelle, and HOL Light \cite{yang2023leandojo,zheng2109minif2f}. At the same time, the broader trustworthy AI community has emphasized that high benchmark performance alone cannot guarantee correctness, safety, or reliability in real-world deployments.

Despite remarkable progress, current reasoning systems continue to exhibit major limitations. LLM-generated reasoning frequently appears mathematically coherent while still containing subtle logical inconsistencies, invalid symbolic transitions, fabricated lemmas, unsupported theorem applications, or semantically incorrect proof steps. In many cases, these failures remain difficult to detect manually because the generated reasoning retains high linguistic fluency and structural plausibility. This distinction between linguistic plausibility and symbolic correctness has emerged as one of the central challenges in modern trustworthy reasoning research.

The problem becomes particularly significant in safety-critical domains such as autonomous robotics, aerospace systems, healthcare, cyber-physical infrastructures, scientific computing, and industrial automation. In such environments, even small reasoning inconsistencies may propagate into unsafe decisions or unreliable system behavior. Consequently, future AI systems will require reasoning pipelines that are not only capable, but also verifiable, explainable, runtime-monitored, and reliability-aware.

Most existing research approaches trustworthy reasoning from isolated perspectives. The theorem-proving community focuses primarily on symbolic proof construction and formal correctness. Runtime verification researchers emphasize behavioral monitoring and execution-level assurance \cite{leucker2009brief}. Neuro-symbolic AI focuses on integrating neural learning with symbolic reasoning \cite{garcez2023neurosymbolic}, while trustworthy AI research emphasizes robustness, explainability, alignment, and reliability \cite{russell2022human}.
However, there is currently no unified operational framework capable of integrating these diverse components into a coherent reasoning lifecycle.

This paper introduces \textsf{ReasonOps}, a new operational paradigm for trustworthy verified reasoning systems. Inspired by DevOps and MLOps, ReasonOps treats reasoning itself as a continuously evolving operational process involving symbolic verification, runtime assurance, uncertainty estimation, adaptive correction, and continual reliability monitoring. Rather than viewing reasoning as a single inference event, ReasonOps models reasoning as a closed-loop operational ecosystem.

\section{Motivation for ReasonOps}

The growing complexity of modern AI reasoning systems has created a widening gap between generative capability and reasoning trustworthiness. Current LLMs excel at producing fluent outputs, yet they frequently lack formal correctness guarantees. Benchmark-level reasoning performance therefore cannot reliably guarantee symbolic validity, semantic consistency, or runtime reliability.

Existing verification approaches remain fragmented and often disconnected from practical deployment pipelines. Interactive theorem provers provide extremely strong correctness guarantees, but they typically operate in isolated offline environments requiring substantial human expertise. Runtime verification systems provide execution-level assurance, yet they rarely integrate deeply with symbolic reasoning architectures. Neuro-symbolic systems improve reasoning structure but still struggle with long-horizon symbolic consistency and uncertainty propagation.

The broader AI ecosystem has historically addressed similar complexity challenges through operational paradigms. DevOps unified software development and deployment into continuous operational workflows, while MLOps later integrated model training, deployment, monitoring, and lifecycle management into scalable machine learning infrastructures. These paradigms demonstrated that increasingly complex intelligent systems require operational coordination mechanisms rather than isolated computational components.
Reasoning systems now face an analogous transition.

Future AI systems will increasingly require continuous reasoning validation, symbolic correctness checking, runtime monitoring, probabilistic trust estimation, adaptive proof repair, behavioral assurance, and system-level reliability management. These requirements suggest the need for a dedicated operational reasoning ecosystem capable of integrating theorem proving, symbolic reasoning, runtime verification, uncertainty estimation, and adaptive correction into unified trustworthy AI pipelines.
ReasonOps addresses this need by introducing an end-to-end operational paradigm for trustworthy reasoning systems.

\section{The ReasonOps Paradigm}

ReasonOps is defined as an operational paradigm for designing, deploying, monitoring, verifying, and continuously maintaining trustworthy AI reasoning systems.
Unlike conventional reasoning pipelines that treat inference as a static generation process, ReasonOps models reasoning as a dynamic operational lifecycle involving multiple interacting verification and assurance layers. The central philosophy behind ReasonOps is that trustworthy reasoning should emerge from continual interaction between neural generation, symbolic reasoning, theorem proving, runtime verification, probabilistic reliability estimation, and adaptive correction mechanisms.

Under this paradigm, reasoning is not considered complete when a response is generated. Instead, reasoning remains subject to continual monitoring, symbolic validation, runtime assurance, and reliability assessment throughout execution. ReasonOps therefore extends reasoning systems beyond traditional theorem proving by incorporating operational trust management directly into the reasoning pipeline.

The paradigm is built upon several foundational principles. First, reasoning should remain formally verifiable whenever possible. Symbolic verification and theorem proving should act as core operational components rather than optional post-processing tools. Second, reasoning systems should support runtime assurance because autonomous AI systems operating continuously in dynamic environments require online monitoring capable of detecting unsafe reasoning trajectories during execution.
Third, reasoning systems should incorporate uncertainty awareness. Since neural reasoning is inherently probabilistic, trustworthy reasoning pipelines must estimate confidence, reliability, and uncertainty propagation. Fourth, reasoning systems should support adaptive correction. Failed proofs, inconsistent symbolic states, and runtime anomalies should trigger repair mechanisms rather than complete reasoning failure. Finally, reasoning systems should remain explainable and auditable so that reasoning traces can be inspected, interpreted, and validated by human operators.
These principles collectively form the conceptual foundation of ReasonOps.


\begin{figure*}[!ht]
\centering
\includegraphics[width=0.95\textwidth]{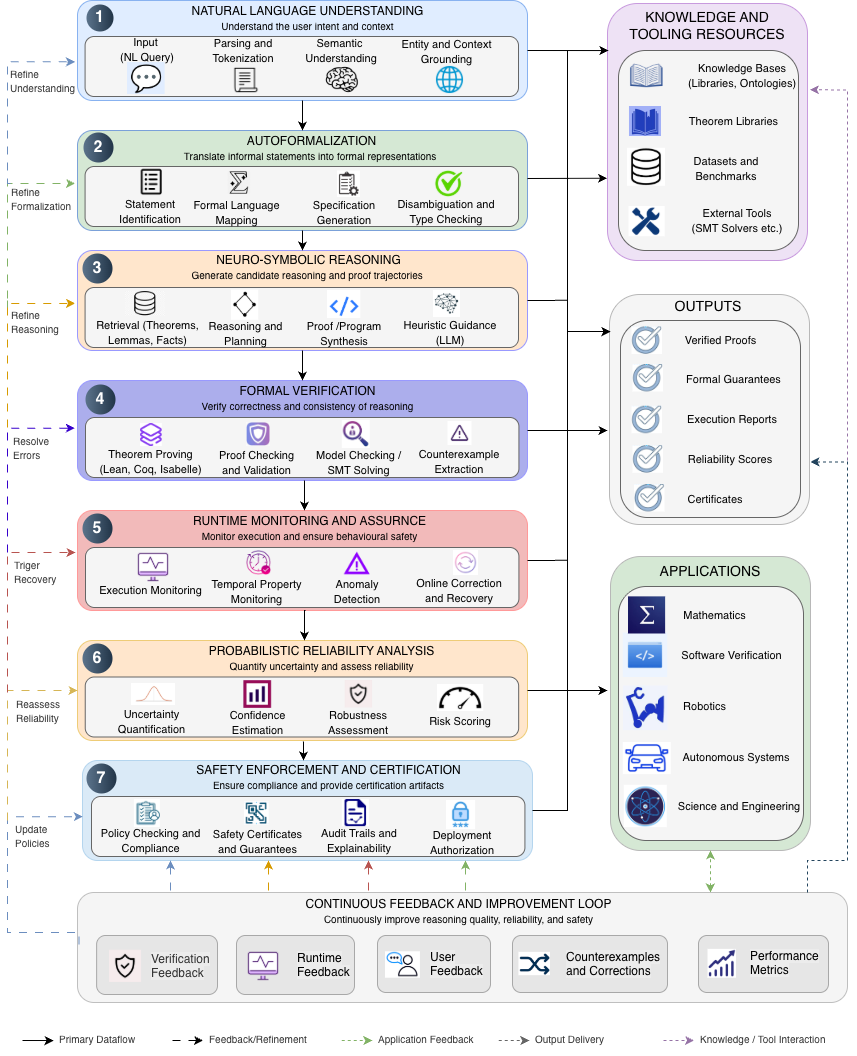}
\caption{ReasonOps Framework}
\label{FIG:reasonops}
\end{figure*}

\section{The ReasonOps Architecture}

The proposed ReasonOps architecture organizes trustworthy reasoning into multiple operational layers, as shown in Figure~\ref{FIG:reasonops}.
The first layer involves semantic understanding and contextual interpretation. Informal reasoning frequently contains ambiguous notation, incomplete assumptions, implicit constraints, and context-dependent semantics. Robust semantic interpretation is therefore necessary before formal reasoning can occur.
The second layer performs autoformalization. At this stage, informal reasoning is transformed into machine-verifiable symbolic representations compatible with theorem provers, SMT solvers, symbolic reasoning engines, and other formal verification systems \cite{wu2022autoformalization}.

The third layer involves symbolic reasoning and proof synthesis. This stage integrates theorem provers, symbolic deduction engines, retrieval-augmented reasoning systems, neuro-symbolic agents, and logical planning frameworks to generate candidate reasoning trajectories \cite{yang2023leandojo}.
The fourth layer performs formal verification and symbolic consistency checking. Generated proofs and reasoning traces are validated against the inference rules of the underlying formal system. Verification engines may involve interactive theorem provers, SMT solvers, temporal logic checkers, or probabilistic verification systems.
The fifth layer introduces runtime assurance and behavioral monitoring. Runtime monitors continually evaluate reasoning trajectories for unsafe transitions, anomalous symbolic states, policy violations, or unstable behavioral patterns during deployment \cite{leucker2009brief}.

The sixth layer incorporates probabilistic reliability estimation. Reliability models evaluate uncertainty propagation, confidence calibration, stochastic correctness, and reasoning stability throughout the reasoning process.
Finally, the seventh layer enforces adaptive correction and continual refinement. When inconsistencies, unsafe states, or low-confidence outputs are detected, reasoning pipelines dynamically repair symbolic states, regenerate proof steps, refine constraints, or update reasoning trajectories.

\section{Operational Workflow of ReasonOps}

The operational workflow of ReasonOps differs significantly from conventional inference pipelines.
Traditional LLM reasoning systems typically follow a relatively simple generation paradigm, i.e.,  $\text{Input} \rightarrow \text{Inference} \rightarrow \text{Output}$.
In contrast, ReasonOps introduces continual interaction between generation, verification, runtime assurance, uncertainty estimation, and adaptive correction.

A typical ReasonOps workflow begins when an informal reasoning task enters the semantic interpretation layer, where contextual assumptions, symbolic entities, latent constraints, and environmental conditions are identified. The task is then autoformalized into symbolic structures compatible with formal reasoning systems.
Next, symbolic reasoning engines and theorem provers generate candidate proofs or reasoning trajectories. These outputs are immediately passed through formal verification layers that validate symbolic correctness and consistency. Simultaneously, runtime assurance mechanisms monitor reasoning stability, detect unsafe symbolic transitions, and identify anomalous behavioral patterns.

Probabilistic trust estimation modules continuously evaluate uncertainty, confidence calibration, reliability propagation, and stochastic correctness throughout the reasoning process. If inconsistencies or unsafe states are detected, adaptive correction mechanisms dynamically repair reasoning trajectories and regenerate symbolic structures.
The resulting workflow transforms reasoning into a closed-loop operational lifecycle rather than a static inference pipeline.


\section{Illustrative Example: Autonomous Braking System Analysis}

To demonstrate the operational workflow of ReasonOps, consider an autonomous braking controller operating in a self-driving vehicle. The system must determine whether emergency braking should be activated when an obstacle is detected under uncertain environmental conditions.

Suppose the following high-level requirement is provided in natural language:

\vspace*{0.1cm}

\begin{mdframed}

\small

\texttt{“Ensure that the vehicle stops safely when an obstacle is detected within 30 meters on a wet road while traveling at 60 km/h.”}

\end{mdframed}

\vspace*{0.2cm}

\normalsize

Traditional LLM-based reasoning systems may generate plausible explanations regarding braking distance, friction, or safe stopping conditions. However, such reasoning may still contain hidden inconsistencies, invalid assumptions, or unsupported symbolic transitions.
Under the proposed ReasonOps paradigm, the reasoning process becomes operationally verifiable.

The natural-language requirement first enters the semantic understanding layer, where contextual entities, environmental assumptions, and safety constraints are extracted. Important variables such as vehicle speed, braking distance, road condition, and stopping constraints are identified. The autoformalization layer then converts the informal requirement into symbolic constraints and formal representations compatible with theorem provers, SMT solvers, or hybrid verification systems.

Next, symbolic reasoning engines compute candidate braking trajectories while incorporating friction models, stopping-distance equations, and environmental uncertainty. Formal verification modules subsequently validate whether safety constraints remain satisfied under all admissible conditions.
Simultaneously, runtime assurance mechanisms monitor live sensor inputs, road conditions, obstacle detection confidence, and dynamic environmental changes. Probabilistic reliability modules estimate confidence levels and evaluate uncertainty propagation across the reasoning process.
If inconsistencies, unsafe states, or low-confidence decisions are detected, adaptive correction modules dynamically refine the reasoning trajectory and regenerate candidate decisions.

In the illustrated example, the system ultimately verifies that braking can safely occur within acceptable safety bounds under wet-road conditions while maintaining a high confidence score.
This example highlights one of the key advantages of ReasonOps, i.e., reasoning is no longer treated as a static inference event, but rather as a continuously monitored and operationally verified lifecycle integrating symbolic correctness, runtime assurance, uncertainty estimation, and adaptive refinement.

\section{Applications of ReasonOps}

ReasonOps has potential applications across multiple domains requiring trustworthy autonomous reasoning.
In autonomous robotics, ReasonOps may provide runtime-verified planning and safety-assured decision pipelines. Symbolic verification layers could validate safety constraints while runtime monitors detect unsafe behavioral trajectories.
Similarly, in healthcare systems, ReasonOps may support reliability-aware diagnostic reasoning and formally verified clinical decision pipelines. Scientific computing systems could employ ReasonOps for theorem-guided scientific discovery, symbolic equation verification, and correctness-aware computational reasoning.

Cyber-physical infrastructures may benefit from runtime-assured reasoning pipelines capable of continually monitoring safety properties during execution. Software engineering workflows could integrate ReasonOps into verified program synthesis, symbolic debugging, formally checked code generation, and correctness-aware software maintenance pipelines.
The paradigm may also become increasingly important for future autonomous AI agents operating continuously in dynamic and uncertain environments.

\section{Open Challenges}

Although ReasonOps provides a conceptual foundation for trustworthy operational reasoning, several important challenges remain unresolved.

One major challenge involves scalability. Large theorem libraries, symbolic search spaces, and runtime monitoring systems may create substantial computational bottlenecks. Another challenge concerns semantic faithfulness because formal correctness alone does not necessarily guarantee preservation of intended semantic meaning during autoformalization and symbolic translation.
Runtime assurance for long-horizon autonomous reasoning also remains comparatively underexplored. Future reasoning systems may require sophisticated behavioral monitoring frameworks capable of detecting unsafe reasoning trajectories over extended operational timescales.

Probabilistic trust estimation presents another difficult research problem because neural reasoning behavior remains fundamentally stochastic and distribution-sensitive. Finally, integrating neural flexibility with symbolic rigor continues to represent one of the central open problems in trustworthy AI research.

\section{Conclusion}

This paper introduced \textit{ReasonOps}, a new operational paradigm for trustworthy verified reasoning systems.
Unlike conventional reasoning pipelines that treat inference as an isolated generation task, ReasonOps models reasoning as a continuously monitored operational lifecycle integrating semantic interpretation, autoformalization, symbolic reasoning, formal verification, runtime assurance, probabilistic reliability estimation, and adaptive correction.
The proposed paradigm extends the operational philosophy of DevOps and MLOps into the domain of AI reasoning by treating reasoning systems as continuously verifiable and reliability-aware infrastructures rather than standalone inference engines.

The paper further presented the layered ReasonOps architecture and demonstrated its operational workflow through an autonomous braking system analysis example. The example illustrated how neural reasoning, symbolic verification, runtime monitoring, and uncertainty estimation can cooperate within a unified reasoning pipeline to support trustworthy autonomous decision making.

As AI systems become increasingly autonomous and reasoning-intensive, the limitations of purely generative reasoning architectures are becoming increasingly evident. Future AI ecosystems will likely require reasoning systems capable not only of producing outputs, but also of continuously justifying, validating, monitoring, repairing, and certifying their reasoning behavior during deployment.
ReasonOps provides an initial conceptual foundation toward this broader vision. In the long term, operational reasoning paradigms such as ReasonOps may become foundational infrastructure for next-generation trustworthy AI systems operating in safety-critical, knowledge-intensive, and dynamically evolving environments.

\bibliographystyle{ACM-Reference-Format}
\bibliography{sample-base}

\end{document}